\documentclass[10pt, final, letterpaper, oneside, twocolumn]{IEEEtran}

\usepackage{cite}
\usepackage{graphicx}
\usepackage{amssymb,amsmath, amsthm}
\usepackage{amsfonts}
\usepackage{multirow}
\usepackage{flushend}
\usepackage{bm}
\usepackage{color}
\newtheorem{remark}{Remark}

\begin{document}

\title{\huge{Impact of Co-Channel Interference on Performance of Multi-Hop Relaying over Nakagami-$m$ Fading Channels}}

\author{Minghua~Xia,~\IEEEmembership{Member,~IEEE}, and Sonia~A\"{\i}ssa,~\IEEEmembership{Senior Member,~IEEE}
\thanks{Manuscript received October 31, 2013. The associate editor coordinating the review of this letter and approving it for publication was M. Morelli.}
\thanks{This work was supported by a Discovery Accelerator Supplement (DAS) Grant from the Natural Sciences and Engineering Research Council (NSERC) of Canada.}
\thanks{The authors are with Institut National de la Recherche Scientifique (INRS), University of Quebec, Montreal, QC, Canada, H5A 1K6 (e-mail: \{xia, aissa\}@emt.inrs.ca).}
\thanks{Digital Object Identifier XXXXXX}
}

\markboth{IEEE WIRELESS COMMUNICATIONS LETTERS, ACCEPTED FOR PUBLICATION} {XIA \MakeLowercase{\textit{et al.}}: Impact of Co-Channel Interference on Performance of Multi-Hop Relaying over Nakagami-$m$ Fading}

\maketitle

\pubid{XXXXXX~\copyright~2013 IEEE}

\begin{abstract}
\noindent This paper studies the impact of co-channel interferences (CCIs) on the system performance of multi-hop amplify-and-forward (AF) relaying, in a simple and explicit way. For generality, the desired channels along consecutive relaying hops and the CCIs at all nodes are subject to Nakagami-$m$ fading with different shape factors. This study reveals that the diversity gain is determined only by the fading shape factor of the desired channels, regardless of the interference and the number of relaying hops. On the other hand, although the coding gain is in general a complex function of various system parameters, if the desired channels are subject to Rayleigh fading, the coding gain is inversely proportional to the accumulated interference at the destination, i.e. the product of the number of relaying hops and the average interference-to-noise ratio, irrespective of the fading distribution of the CCIs.
\end{abstract}

\section{Introduction} \label{Section:Introduction}
Multi-hop relaying technique is a cost-effective solution to the pressing problem of how to extend wireless coverage and mitigate channel impairments without extra transmit power. The main idea behind multi-hop relaying is to exploit multiple intermediate idle nodes between a source node and its far-end destination to forward signals in a consecutive way. Among different schemes for the relaying, this letter focus on the amplify-and-forward (AF) technique due to its low complexity for practical implementation.

\IEEEpubidadjcol

When multi-hop AF relaying is practically deployed, its achievable performance is inevitably degraded by co-channel interferences (CCIs), coming from external interfering sources. In order to study the effect of CCIs on the system performance, in \cite{Soithong11CL08, IkkiTVT12Feb} the CCIs were assumed to be subject to Rayleigh fading and the system performance was extensively explored, in terms of outage probability, average symbol error probability (ASEP) and ergodic capacity. Recently, assuming CCIs to be subject to Nakagami-$m$ fading, the outage probability, the ASEP and the ergodic capacity were analyzed in \cite{Soithong12TVT03, Wen12CL09, Trigui13TCOM07}, respectively. Actually, if CCIs are subject to Nakagami-$m$ fading, due to extreme difficulty of mathematical tractability, the aforementioned performance metrics cannot be uniformly analyzed but have to be dealt with separately. Even so, the obtained results in \cite{Soithong12TVT03, Wen12CL09, Trigui13TCOM07} are generally too complex to offer insights, in a simple and explicit way, as to what parameters determine system performance.

In this paper, we investigate multi-hop cooperative AF relaying system in the presence of CCIs subject to Nakagami-$m$ fading. Our purpose here being to gain insights into the system performance in a simple and explicit way, we particularly investigate the impact of Nakagami-$m$ faded interferences on the system. To this end, we exploit lower and upper bounds on the end-to-end signal-to-interference ratio (SIR) of multi-hop AF relaying transmission and derive their distribution functions. Afterwards, the resultant distribution functions are applied to analyze the diversity gain and the coding gain of the considered system, which are two key indicators of outage probability and ASEP at high signal-to-noise ratio (SNR).

Our results disclose that the {\it diversity gain} of multi-hop AF transmission in the presence of CCIs is determined only by the fading shape factor of the desired channels, regardless of the CCIs and the number of relaying hops. On the other hand, the {\it coding gain} is explicitly obtained in terms of various system parameters. In particular, if the desired channels are subject to Rayleigh fading, the coding gain is shown to be  inversely proportional to the accumulated interference at the destination, i.e. the product of the number of relaying hops and the average interference-to-noise ratio, irrespective of the fading distribution of the CCIs. This is the first time to disclose the effect of interference accumulation in multi-hop relaying impaired by interferences, in parallel to the well-known effect of noise accumulation in multi-hop relaying affected only by thermal noises \cite{BoradeTIT07Oct}.

In detailing the above highlighted contributions, the rest of the letter is organized as follows. Section~\ref{Section:SystemModel} details the system model. Then, the distribution functions of the lower and upper bounds on the end-to-end SIR are developed in Section~\ref{Section:BoundsSIR}. Subsequently, the system performance is extensively analyzed and discussed in Section~\ref{PerformanceAnalyses} and, finally, concluding remarks are provided in Section~\ref{Section:Conclusion}.

\IEEEpubidadjcol
\section{The System Model}  \label{Section:SystemModel}
\begin{figure}[t]
\centering
\includegraphics [width=3.2in, clip, keepaspectratio]{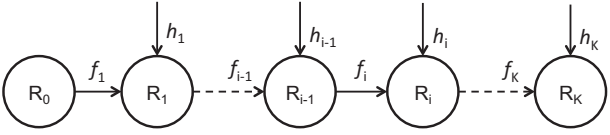}
\caption{System model of multi-hop cooperative amplify-and-forward transmission in the presence of co-channel interferences. The desired channels $f_i$ and the interfering channels $h_i, \forall i \in [1, K]$, are subject to Nakagami fading with different shape parameters.}  \label{Fig.SystemModel}
\end{figure}

Figure~\ref{Fig.SystemModel} illustrates a generic model of multi-hop cooperative AF relaying system in the presence of CCIs, where the source node $\mathrm{R}_0$ communicates with its destination $\mathrm{R}_K$ through the assistance of $K-1$ consecutive AF relays.  It is assumed that all nodes in the system are equipped with a single half-duplex omnidirectional antenna each. All nodes work in a time-division multiple access (TDMA) fashion and equal time slots of a transmission frame are allocated to each node. Moreover, only one node transmits to its next neighbor along the multi-hop link during each time slot, with a fixed transmit power $P$. Each relaying node along the multi-hop path and the destination node are interfered by different external interfering sources (cf. $h_i$, $\forall i \in [1, K]$, in Fig.~\ref{Fig.SystemModel}).

For generality of subsequent performance analysis, the desired channel $f_i$ of the $i^\mathrm{th}$-hop (i.e. the link from $\mathrm{R}_{i-1}$ to $\mathrm{R}_i$ as shown in Fig.~\ref{Fig.SystemModel}) and the CCI $h_i$ at $\mathrm{R}_i$, $\forall i \in [1, K]$, are assumed to be subject to Nakagami-$m$ fading with shape factors $\alpha_i$ and $\beta_i$, respectively. Specifically, the probability density function (PDF) of $|f_i|^2$ can be explicitly given by
\begin{equation} \label{Eq.PDFchannel}
f_{|f_i|^2}(x) = \frac{\alpha_i}{\Gamma{(\alpha_i)\,\bar{\gamma}_{_{Di}}}}\left(\frac{\alpha_i}{\bar{\gamma}_{_{Di}}}\,x\right)^{\alpha_i - 1}
\exp\left(-\frac{\alpha_i}{\bar{\gamma}_{_{Di}}}\,x\right)
\end{equation}
where $\Gamma(x) = \int_0^\infty{t^{x-1}e^{-t}\,\mathrm{d}t}$, $\forall x > 0$, denotes the Gamma function, $\bar{\gamma}_{_{Di}}$ denotes the average SNR of the desired channel at the $i^\mathrm{th}$ hop. The PDF of $|h_i|^2$ can be defined like \eqref{Eq.PDFchannel}, yet with the parameters $(\alpha_i, \bar{\gamma}_{_{Di}})$ replaced by $(\beta_i, \bar{\gamma}_{_{Ii}})$, where $\bar{\gamma}_{_{Ii}}$ stands for the average interference-to-noise ratio (INR) of the CCI at the $i^\mathrm{th}$ relay.

Then, the received SIR at the $i^\mathrm{th}$ relaying hop is shown as:
\begin{equation} \label{Eq.SIR}
\gamma_i = \frac{P|f_i|^2}{|h_i|^2},
\end{equation}
where the thermal noise at the relay $\mathrm{R}_i$ is ignored since the system under study is assumed  interference-limited. Furthermore, using a similar procedure to that in \cite{IkkiTVT12Feb}, the end-to-end SIR from the source node, $\mathrm{R}_0$, to its final destination, $\mathrm{R}_K$, can be easily shown to be given by
\begin{equation} \label{Eq.e2eSIR}
\gamma_{e2e} = \left(\sum_{i=1}^{K}\frac{1}{\gamma_i}\right)^{-1}.
\end{equation}

\begin{remark}[Number of interfering signals]
In this paper, we consider that each relaying node or the final destination is affected by a single interferer. Actually, if $N > 1$ independent and identically Nakagami-$m$ faded interferes are considered, they can be exactly expressed by a single Nakagami-$m$ faded interferer with shape factor $mN$, by recalling the fact that the sum of multiple independent and identically Gamma-distributed random variables is still Gamma-distributed. Also, if $N > 1$ independent but non-identically Nakagami-$m$ faded interferes are assumed, they are still equivalent to a single Nakagami-$m$ faded interferer by using the moment matching method, see e.g. \cite{Costa12TVT08, Xia12JSAC09}.
\end{remark}
\section{The Lower and Upper Bounds on the End-to-End SIR} \label{Section:BoundsSIR}
In order to facilitate the mathematical tractability on the statistics of the end-to-end SIR  $\gamma_{e2e}$ given by \eqref{Eq.e2eSIR}, we investigate the lower and upper bounds on $\gamma_{e2e}$. These bounds facilitate, otherwise extremely tedious, mathematical manipulations \cite{Soithong12TVT03, Wen12CL09, Trigui13TCOM07}. More importantly, they enable gaining illuminating insights into system performance in terms of the diversity and coding gains, as shown in the next section. Specifically, in light of \eqref{Eq.e2eSIR}, it follows that the end-to-end SIR is upper bounded by
\begin{equation} \label{Eq.e2eSIRUpperBound}
\gamma_{e2e} \le \left(\max_{i=1, \cdots, K}\frac{1}{\gamma_i}\right)^{-1} = \min_{i=1, \cdots, K}{\gamma_i}.
\end{equation}
On the other hand, it is clear that $\gamma_{e2e}$ is lower bounded by
\begin{equation} \label{Eq.e2eSIRLowerBound}
\gamma_{e2e} \ge \left(K \max_{i=1, \cdots, K}\frac{1}{\gamma_i}\right)^{-1} = \frac{1}{K}\min_{i=1, \cdots, K}{\gamma_i}.
\end{equation}
As a consequence, combining \eqref{Eq.e2eSIRUpperBound} and \eqref{Eq.e2eSIRLowerBound} yields the lower and upper bounds on $\gamma_{e2e}$, namely,
\begin{equation} \label{Eq.e2eSIRBounds}
\underbrace{\frac{1}{K}\min_{i=1, \cdots, K}{\gamma_i}}_{\gamma_{e2e}^\mathrm{lower}}
\le \gamma_{e2e}
\le  \underbrace{\min_{i=1, \cdots, K}{\gamma_i}}_{\gamma_{e2e}^\mathrm{upper}}.
\end{equation}
Moreover, it is observed from \eqref{Eq.e2eSIRBounds} that
\begin{equation} \label{Eq.e2eSIRBoundsRelation}
\gamma_{e2e}^\mathrm{lower}
= \frac{1}{K}\,\gamma_{e2e}^\mathrm{upper}.
\end{equation}

In the sequel, we derive the distribution function of the upper bound $\gamma_{e2e}^{\mathrm{upper}}$. That of the lower bound $\gamma_{e2e}^{\mathrm{lower}}$ follows in a straightforward manner. To this end, we first need the PDF of $V \triangleq |f_k|^2/|h_k|^2$. By virtue of \eqref{Eq.PDFchannel}, it is clear that $V$ is the ratio of two Gamma-distributed variables and, thus, its PDF is of the beta distribution of the second kind, given by
\begin{equation} \label{Eq.PDF-V}
f_{V}(x)
= \frac{a_1}{B(\alpha_i, \beta_i)} (a_1 x)^{\alpha_i-1}(1+a_1 x)^{-(\alpha_i + \beta_i)},
\end{equation}
where $B(p, q) = \int_0^1{x^{p-1}(1-x)^{q-1}}\,\mathrm{d}x$, $\forall p,\,q > 0$, denotes the beta function, and $a_1 \triangleq \alpha_i \, \bar{\gamma}_{_{Ii}}/(\beta_i \, \bar{\gamma}_{_{Di}})$. Then, by \eqref{Eq.PDF-V}, the PDF of $\gamma_i = PV$ as shown in \eqref{Eq.SIR} can be easily expressed as
\begin{equation} \label{Eq.PDF-SIR@Rk}
f_{\gamma_i}(x)
= \frac{a_2}{B(\alpha_i, \beta_i)} (a_2 x)^{\alpha_i-1}(1+a_2 x)^{-(\alpha_i + \beta_i)},
\end{equation}
where $a_2 \triangleq \alpha_i \, \bar{\gamma}_{_{Ii}}/(P \beta_i \, \bar{\gamma}_{_{Di}})$. Then, by integrating $f_{\gamma_i}(\gamma)$ shown in \eqref{Eq.PDF-SIR@Rk} with respect to $x$ and appealing to \cite[Eq. (3.194.1)]{Gradshteyn07}, we obtain the CDF of $\gamma_i$, namely,
\begin{equation} \label{Eq.CDF-SIR@Rk}
F_{\gamma_i}(x)
= \frac{(a_2 x)^{\alpha_i}}{\alpha_i B(\alpha_i, \beta_i)} \, {_2F_1}\left(\alpha_i, \alpha_i+\beta_i; \alpha_i+1; -a_2 x\right),
\end{equation}
where ${_2F_1}(a, b; c; x)$ is the Gaussian hypergeometric function \cite[Eq. (9.100)]{Gradshteyn07}.

With the resultant \eqref{Eq.CDF-SIR@Rk} and by recalling the theory of order statistics, the CDF of the upper bound on the end-to-end SIR shown in \eqref{Eq.e2eSIRBounds} can be expressed as
\begin{equation}  \label{Eq.CDF-SIR-Upper}
F_{\gamma_{e2e}^{\mathrm{upper}}}(x)
= 1-\prod_{i=1}^K{\left[1-F_{\gamma_i}(x)\right]}.
\end{equation}

\textbf{Case I (Symmetric Channels):} If the desired channels along consecutive relaying hops are symmetric (i.e. $\alpha_i \triangleq \alpha$ and $\bar{\gamma}_{_{Di}} \triangleq \bar{\gamma}_{_{D}}$, $\forall i \in [1, K]$) and the interfering channels at all nodes are symmetric as well (i.e.  $\beta_i \triangleq \beta$ and $\bar{\gamma}_{_{Ii}} \triangleq \bar{\gamma}_{_{I}}$, $\forall i \in [1, K]$), the CDF in \eqref{Eq.CDF-SIR-Upper} reduces to
\begin{equation}   \label{Eq.CDF-SIR-Upper-2}
F_{\gamma_{e2e}^{\mathrm{upper}}}(\gamma) = 1 - [1-F_{\gamma_{\mathrm{sym}}}(x)]^{K},
\end{equation}
where $F_{\gamma_{\mathrm{sym}}}(x) = \frac{(a_3 x)^{\alpha}}{\alpha \,B(\alpha, \beta)} \, {_2F_1}\left(\alpha, \alpha+\beta; \alpha+1; -a_3 x\right)$ with $a_3 \triangleq \alpha \, \bar{\gamma}_{_{I}}/(P \beta \, \bar{\gamma}_{_{D}})$.

\textbf{Case II (Non-Symmetric Channels):} In reality, of course, the symmetry assumed in the above Case I may not always be satisfied. In such a case, by recalling the fact that the end-to-end SIR is dominated by the worst hop among $K$ consecutive hops, as implied by \eqref{Eq.e2eSIRBounds}, we may choose the worst hop as a benchmark and set the parameters of other hops to be identical to those of this worst hop so as to obtain an approximated expression of \eqref{Eq.CDF-SIR-Upper}. Specifically, the worst hop can be identified as follows. Since the received SIR at $i^\mathrm{th}$-hop (i.e. $\gamma_i$ as shown in \eqref{Eq.SIR}) is of the beta distribution of the second kind as shown in \eqref{Eq.PDF-SIR@Rk}, it is not hard to show that the mean of $\gamma_i$ is given by
\begin{equation}  \label{AverageSIR}
\bar{\gamma}_i = \frac{\alpha_i}{a_2 (\beta_i-1)}
= \frac{P \beta_i \bar{\gamma}_{_{Di}}}{(\beta_i - 1) \bar{\gamma}_{_{Ii}}}.
\end{equation}
As a consequence, by using the minimum average SIR criterion, the worst hop among $K$ consecutive relaying hops can be explicitly determined by
\begin{equation}  \label{WorstHop}
\hat{i} = \arg\min_{i=1, \cdots, K}{\frac{P \beta_i \bar{\gamma}_{_{Di}}}{(\beta_i - 1) \bar{\gamma}_{_{Ii}}}}.
\end{equation}
Accordingly, \eqref{Eq.CDF-SIR-Upper} can be approximated by
\begin{equation}  \label{Eq.CDF-SIR-Upper-3}
F_{\gamma_{e2e}^{\mathrm{upper}}}(x)
\approx 1-{\left[1-F_{\gamma_{\hat{i}}}(x)\right]^K}.
\end{equation}

With the resultant CDF of the upper bound given by \eqref{Eq.CDF-SIR-Upper}, \eqref{Eq.CDF-SIR-Upper-2} or \eqref{Eq.CDF-SIR-Upper-3} and in light of the relation shown in \eqref{Eq.e2eSIRBoundsRelation}, it is clear that the CDF of the lower bound $\gamma_{e2e}^{\mathrm{lower}}$ can be expressed as
\begin{equation}\label{Eq.CDF-SIR-Lower}
F_{\gamma_{e2e}^{\mathrm{lower}}}(\gamma)
= F_{\gamma_{e2e}^{\mathrm{upper}}}(Kx).
\end{equation}
\section{Performance Analyses and Discussions}
\label{PerformanceAnalyses}
\subsection{Outage Probability}  \label{Section:Outage}
With the resultant CDFs of the lower and upper bounds on the end-to-end SIR and by recalling the fact that outage probability is a monotonically decreasing function of the end-to-end SIR, the outage probability can be readily shown to be bounded by
\begin{equation} \label{Eq.OutageProbability}
F_{\gamma_{e2e}^{\mathrm{upper}}}\left(\gamma_{_\mathrm{th}}\right)
\le \mathrm{P_{outage}}(\gamma_{_\mathrm{th}})
\le F_{\gamma_{e2e}^{\mathrm{lower}}}\left(\gamma_{_\mathrm{th}}\right).
\end{equation}

In order to confirm the effectiveness of the lower and upper bounds in \eqref{Eq.OutageProbability}, extensive simulation experiments were performed to compare the simulation results of outage probability with the numerical ones computed by \eqref{Eq.OutageProbability}. In the Monte-Carlo simulations conducted, all desired channels along the multi-hop relaying link were subject to Nakagami-$m$ fading with the same shape factor $\alpha$ while the CCIs at each node were subject to similar fading but with shape factor $\beta$. The variance of thermal noise at each node was set to unity; the transmit power and the INR at each node were defined in the unit of dB with respect to the noise variance.

\begin{figure}[t]
\centering
\includegraphics [width=2.9in, clip, keepaspectratio]{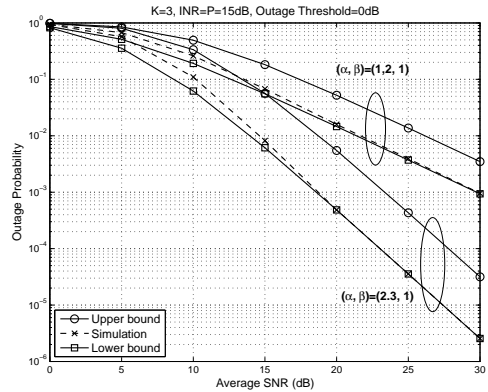}
\caption{Outage probability of a $3$-hop AF relaying with different fading shape factors $\alpha$, under a given $\beta=1$.}  \label{Fig.Outage_1}
\end{figure}

Figure~\ref{Fig.Outage_1} shows the outage probability of a $3$-hop AF relaying in the presence of CCIs, where the value of the fading shape factor of the CCIs was fixed to $1$ (i.e. $\beta=1$) whereas the fading parameter of the desired channels along the relaying link was varied ($\alpha=1.2, 2.3$). It is observed that, for any given values of $(\alpha, \beta)$, the simulation results are very tight with  the lower bound, particularly in the medium and high SNR regions. On the other hand, it is seen that, for a fixed value of $\beta$, increasing the value of $\alpha$ improves the outage probability significantly, as expected.

\begin{figure}[t]
\centering
\includegraphics [width=2.9in, clip, keepaspectratio]{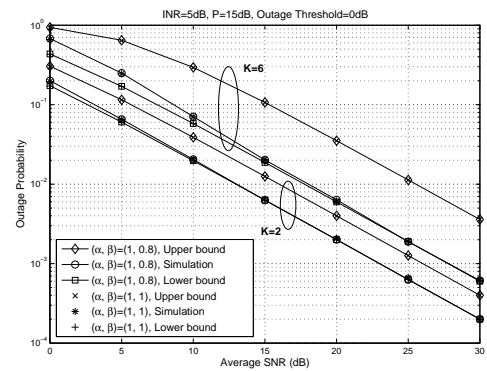}
\caption{Outage probability of a $K$-hop AF relaying $(K=2, 6)$ with different fading shape factors $\beta$, under a given $\alpha=1$.}  \label{Fig.Outage_2}
\end{figure}

Figure~\ref{Fig.Outage_2} shows the outage probability of a $K$-hop AF relaying $(K=2, 6)$ in the presence of CCIs, where the value of $\alpha$ was fixed to $1$ whereas the value of $\beta$ changed from $0.8$ to $1$. It is shown that, for either $K=2$ or $K=6$, the simulation results are still very tight with the lower bound in the medium and high SNR ranges. Moreover, we can see that increasing the number of relaying hops (i.e. $K$) deteriorates the outage probability. On the other hand, for a fixed value of $K$, one can get surprised to find that changing the value of $\beta$ (i.e. the fading shape factor of the CCIs) does not change the outage probability (either the upper bound, the lower bound, or the simulation results) at all. The fundamental reasons behind these observations will be revealed in the next section.

\subsection{Diversity and Coding Gains} \label{Section:DiversityCodingGains}
Here, we derive the diversity gain and the coding gain of the multi-hop AF relaying in the presence of CCIs, which are two key indicators of outage probability and ASEP at high SNR. According to \cite{WangTCOM03Aug}, if the limit of the PDF of the end-to-end SIR can be expressed as
\begin{equation}   \label{Eq.WangI}
\lim\limits_{x \to 0}f_{\gamma_{e2e}}(x) = bx^t+\emph{o}\left(x^{t+\epsilon}\right),
\end{equation}
where $b, \epsilon > 0$ and $\emph{o}(\,.\,)$ pertain to the Landau notation defined as $\lim_{x \to 0}|\emph{o}(x)/x|=0$, then the diversity and coding gains are given by
\begin{equation}  \label{Eq.WangII}
G_d = t+1
\end{equation}
and
\begin{equation}  \label{Eq.WangIII}
G_c = l\left(\frac{2^{t}\,b\,\Gamma(t+\frac{3}{2})}{\sqrt{\pi}(t+1)}\right)^{-\frac{1}{t+1}},
\end{equation}
respectively, where $l$ is a positive constant pertaining to the used constellation, e.g. $l=2$ for binary phase-shift keying.

Now, it is ready to derive the diversity and coding gains of the system under study. Since \eqref{Eq.CDF-SIR-Upper-3} has a similar form to \eqref{Eq.CDF-SIR-Upper-2}, in the following we focus on the symmetric case. The non-symmetric case can be analyzed in a similar way.

At first, differentiating the CDF \eqref{Eq.CDF-SIR-Upper-2} of the upper bound on the end-to-end SIR with respect to $x$ yields its PDF:
\begin{equation} \label{Eq.PDF-SIR-Upper}
f_{\gamma_{e2e}^\mathrm{upper}}(x)
= K f_{\gamma_{\mathrm{sym}}}(x) [1-F_{\gamma_{\mathrm{sym}}}(x)]^{K-1},
\end{equation}
where $f_{\gamma_{\mathrm{sym}}}(x)$ denotes the differential of $F_{\gamma_{\mathrm{sym}}}(x)$ with respect to $x$, which, in view of $F_{\gamma_{\mathrm{sym}}}(x)$ given immediately after \eqref{Eq.CDF-SIR-Upper-2} and the differential relation from \eqref{Eq.CDF-SIR@Rk} to \eqref{Eq.PDF-SIR@Rk}, can be readily expressed as
\begin{equation}  \label{PDF_Symmetric}
f_{\gamma_{\mathrm{sym}}}(x)
= \frac{a_3}{B(\alpha, \beta)} (a_3 x)^{\alpha - 1}(1+a_3 x)^{-(\alpha + \beta)}.
\end{equation}
Then,  inserting \eqref{PDF_Symmetric} into \eqref{Eq.PDF-SIR-Upper} and taking limit as $x \to 0$ yields
\begin{equation}
\lim\limits_{x \to 0}f_{\gamma_{e2e}^\mathrm{upper}}(x)
=  \frac{K a_3^{\alpha}}{B(\alpha, \beta)}x^{\alpha - 1} + \emph{o}\left(x^{\alpha - 1}\right).                                \label{Eq.DiversityCodingGains-B3}
\end{equation}
Afterwards, comparing \eqref{Eq.DiversityCodingGains-B3} with \eqref{Eq.WangI} leads to $t=\alpha-1$ and $b=K a_3^{\alpha}/B(\alpha, \beta)$. As a result, substituting them into \eqref{Eq.WangII}-\eqref{Eq.WangIII} and performing some algebraic manipulations, we obtain
\begin{equation}  \label{Eq.DiversityCodingGains-B4}
G_d = \alpha,
\end{equation}
\begin{equation}  \label{Eq.DiversityCodingGains-B5}
G_c = \frac{lP\beta \bar{\gamma}_D}{\alpha \bar{\gamma}_I} \left(\frac{\sqrt{\pi}\,\alpha \,B(\alpha, \beta)}{2^{\alpha-1}K \, \Gamma(\alpha+\frac{1}{2})}\right)^{\frac{1}{\alpha}}.
\end{equation}

Equation \eqref{Eq.DiversityCodingGains-B4} implies that, for multi-hop AF relaying in the presence of CCIs, the achievable {\it diversity gain} is determined only by the fading shape factor of the desired channels, regardless of the CCIs and the number of relaying hops. This is in agreement with the observation in Fig.~\ref{Fig.Outage_1} where the slopes of all curves increase with the value of $\alpha$, and the observation in Fig.~\ref{Fig.Outage_2} where the slopes of all curves remains unvaried even though the values of $\beta$ and $K$ change.

As for the {\it coding gain}, given by \eqref{Eq.DiversityCodingGains-B5}, it is not hard to show that, in general, $G_c$ decreases monotonically with $\alpha$, $\beta$, $\bar{\gamma}_I$ and $K$ yet increases monotonically with $\bar{\gamma}_D$. That is, larger number of relaying hops ($K$) and stronger CCIs ($\beta$, $\bar{\gamma}_{_{I}}$) deteriorate the coding gain dramatically.

\textbf{Special Case I ($\boldsymbol{\alpha = 1}$):} When the desired channels along consecutive relaying hops are assumed to be subject to Rayleigh fading, setting $\alpha = 1$ in \eqref{Eq.DiversityCodingGains-B4} and \eqref{Eq.DiversityCodingGains-B5} and making some algebraic manipulations yields
\begin{equation}  \label{Eq.DiversityCodingGains-B6}
G_d = 1,
\end{equation}
\begin{equation}  \label{Eq.DiversityCodingGains-B7}
G_c = \frac{2lP\bar{\gamma}_D}{K\bar{\gamma}_I}.
\end{equation}
Equation \eqref{Eq.DiversityCodingGains-B7} demonstrates that the diversity gain is independent of the fading shape factor $\beta$ of the CCIs but is inversely proportional to the product of $K$ and $\bar{\gamma}_I$. This is {\it the effect of interference accumulation} in multi-hop AF relaying when the CCIs are taken into account, just like the well-known effect of noise accumulation originally revealed in \cite{BoradeTIT07Oct} where no CCIs but only thermal noises were considered. Equation \eqref{Eq.DiversityCodingGains-B7} gives explicitly the reason why the outage probability is independent of the values of $\beta$, as previously observed from Fig.~\ref{Fig.Outage_2}.

\textbf{Special Case II ($\boldsymbol{\beta = 1}$):} When the CCIs at the relaying nodes and at the final destination are subject to Rayleigh~fading (corresponding also to the case where each node is interfered by infinite number of interfering signals, by recalling the law of large numbers), setting $\beta = 1$ into \eqref{Eq.DiversityCodingGains-B5} yields
\begin{equation}  \label{Eq.DiversityCodingGains-B8}
G_c = \frac{lP \bar{\gamma}_D}{\alpha \bar{\gamma}_I} \left(\frac{\sqrt{\pi}}{2^{\alpha-1}K \, \Gamma(\alpha+\frac{1}{2})}\right)^{\frac{1}{\alpha}},
\end{equation}
which clearly shows that larger $\alpha$ deteriorates the coding gain.

\begin{remark}[The ASEP]
With the obtained distribution functions of the tight upper bound on the end-to-end SIR, the ASEP can be readily derived by using the PDF/CDF-based methodology. However, since ASEP behaves like outage probability and both of them are characterized by the above diversity and coding gains, we do not discuss ASEP in this letter. For the interested reader, please refer to \cite{WangTCOM03Aug}.
\end{remark}
\section{Conclusion}   \label{Section:Conclusion}
This paper provided a general performance analysis of multi-hop AF relaying in the presence of CCIs, by using a simple but effective upper-bound on the end-to-end SIR. In particular, the impact of external Nakagami-$m$ faded interferences on system performance of multi-hop AF relaying was analytically identified, by means of the diversity gain and the coding gain. Also, the effect of interference accumulation was firstly disclosed in this paper, and found to be in accordance with the well-known effect of noise accumulation in multi-hop relaying affected only by thermal noises.

\vfill

\begin{thebibliography}{99}

\bibitem{Soithong11CL08}
T. Soithong, V. A. Aalo, G. P. Efthymoglou, and C. Chayawan, ``Performance of multihop relay systems with co-channel interference in Rayleigh fading channels,'' \emph{IEEE Commun. Lett.}, vol.~15, no.~8, pp.~836--838, Aug.~2011.

\bibitem{IkkiTVT12Feb}
S.~Ikki and S.~A\"{i}ssa, ``Multihop wireless relaying systems in the presence of cochannel interferences: performance analysis and design optimization,'' \emph{IEEE Trans. Veh. Technol.}, vol.~61, no.~2, pp.~566--573, Feb.~2012.

\bibitem{Soithong12TVT03}
T. Soithong, V. A. Aalo, G. P. Efthymoglou, and C. Chayawan, ``Outage analysis of multihop relay systems in interference-limited Nakagami-$m$ fading channels,'' \emph{IEEE Trans. Veh. Technol.}, vol.~61, no.~3, pp.~1451--1457, Mar.~2012.

\bibitem{Wen12CL09}
M. Wen, X. Cheng, A. Huang, and B. Jiao, ``Asymptotic performance analysis of multihop relaying with co-channel interference in Nakagami-$m$ fading channels,'' \emph{IEEE Commun. Lett.}, vol.~16, no.~9, pp.~1450--1453, Sep.~2012.

\bibitem{Trigui13TCOM07}
I. Trigui, S. Affes, and A. St\'{e}phenne, ``Ergodic capacity analysis for interference-limited AF multi-hop relaying channels in Nakagami-$m$ fading,'' \emph{IEEE Trans. Commun.}, vol.~61, no.~7, pp.~2726--2734, Jul.~2013.

\bibitem{BoradeTIT07Oct}
S.~Borade, L.~Zheng, and R.~Gallager, ``Amplify-and-forward in wireless relay networks: rate, diversity, and network size,'' \emph{IEEE Trans. Inf. Theory}, vol.~53, no.~10, pp.~3302--3318, Oct.~2007.

\bibitem{Costa12TVT08}
D. B. da Costa, H. Ding, M. D. Yacoub, and J. Ge, ``Two-way relaying in interference-limited AF cooperative networks over Nakagami-$m$ fading,''  \emph{IEEE Trans. Veh. Technol.}, vol.~61, no.~8, pp.~3766--3771, Oct. 2012.

\bibitem{Xia12JSAC09}
M.~Xia and S.~A\"{i}ssa, ``Moments based framework for performance analysis of one-way/two-way CSI-assisted AF relaying,'' \emph{IEEE J. Select. Areas Commun.}, vol.~30, no.~8, pp.~1464--1476, Sep.~2012.

\bibitem{Gradshteyn07}
I.~S.~Gradshteyn and I.~M.~Ryzhik, \emph{Table of Integrals, Series and Products}, 7th Ed., Academic Press, 2007.

\bibitem{WangTCOM03Aug}
Z.~Wang and G.~B.~Giannakis, ``A simple and general parameterization quantifying performance in fading channels,'' \emph{IEEE Trans. Commun.}, vol.~51, no.~8, pp.~1389-1398, Aug.~2003.

\end{thebibliography}
\end{document}